\newcommand{\be}{\begin{equation}}
\newcommand{\ee}{\end{equation}}
\newcommand{\ba}{\begin{eqnarray}}
\newcommand{\ea}{\end{eqnarray}}
\renewcommand{\(}{\left(}
\renewcommand{\)}{\right)}

\newcommand{\lk}{\left[}
\newcommand{\rk}{\right]}

\newcommand{\hw}{\hat{\omega}}

\newcommand{\g}{\gamma}
\newcommand{\hq}{\hat{q}}

\input{aipcheck}

\documentclass[
    ,final            
  ]
  {aipproc}

  \usepackage{bbm}
\usepackage{amsmath}
  
\layoutstyle{8x11single}

\begin{document}

\title{Holographic equilibration at strong and intermediate coupling}


\author{Aleksi Vuorinen}{
  address={Helsinki Institute of Physics and Department of Physics, P.O. Box 64, FI-00014 University of Helsinki}
}

\begin{abstract}
In these conference proceedings, I describe recent developments in the study of thermalization dynamics in strongly, but not infinitely strongly coupled field theories using holography. After reviewing the main tools required in these calculations, I introduce a set of central results, discuss their physical implications, and finally outline a number of challenges to be tackled in the future. 
\end{abstract}

\maketitle


\section{Introduction}

During the past 10 years or so, holography has become one of the standard tools to study equilibration dynamics in strongly coupled field theory systems ranging from heavy ion collisions to various condensed matter setups. The reason for this stems most importantly from the difficult nature of the problem when tackled with standard field theory tools: While perturbative techniques typically fail already at moderate couplings, nonperturbative techniques based on lattice Monte Carlo simulations are completely inapplicable for studying time dependent quantum phenomena. At the same time, a set of recent remarkable advances in numerical relativity in AdS space has led to the gauge/gravity duality offering solutions to a range of exceedingly complicated problems with phenomenological applications e.g.~in the study of a thermalizing quark gluon plasma (for reviews, see e.g.~\cite{Gubser:2009md,CasalderreySolana:2011us,Brambilla:2014jmp,Chesler:2013lia}).

Whether insights derived using the gauge/gravity duality remain on the qualitative level or can be elevated to quantitative predictions for heavy ion or condensed matter physics problems depends crucially on whether the duality can be generalized to directly apply to the relevant setups. Achieving this goal would on one hand require constructing initial states closer to the physical ones, and on the other hand relaxing the assumptions of infinite $N_c$, large 't Hooft coupling $\lambda$ as well as conformal invariance inherent in typical holographic studies of ${\mathcal N}=4$ Super Yang-Mills (SYM) theory. Considerable progress has already been achieved on both of these fronts, including e.g.~the construction of colliding shock wave solutions in strongly coupled SYM theory \cite{Chesler:2010bi,Wu:2011yd,Heller:2012km}, the determination of so-called $\alpha'$ corrections to studies of holographic thermalization \cite{Steineder:2012si,Steineder:2013ana,Baron:2013cya,Stricker:2013lma}, as well as the first ever thermalization calculations in non-conformal backgrounds \cite{Craps:2013iaa}.

In the conference proceedings at hand, my goal is to describe some recent developments in one of the above directions, namely including $\alpha'$ corrections to the type IIB supergravity (SUGRA) action when performing calculations in holographic thermalization, thereby taking them away from the $\lambda=\infty$ limit. The results reported here are all for a set of two-point functions, both in and out of thermal equilibrium. The latter involve a simple model of holographic equilibration, where the field theory process is dual to the gravitational collapse of a planar shell in AdS$_5$ spacetime \cite{Danielsson:1999zt}. The obtained Green's functions allow one to study not only the gravitational dynamics with and without the $\alpha'$ corrections, but also the pattern, with which excitations of different energy (length) scales thermalize. As has been argued in \cite{Steineder:2012si,Steineder:2013ana}, at least for some observables lowering the coupling can be seen to have important qualitative effects on the thermalization pattern of the system.

\section{Setup}

The way to proceed beyond the usual $\lambda=\infty$ limit in holographic calculations within the ${\mathcal N}=4$ SYM theory is to account for all effets of order $\alpha'^3$ originating from the type IIB supergravity action. These are typically dressed in the form of an expansion in $\gamma\equiv\frac{1}{8}\zeta(3)\lambda^{-\frac{3}{2}}$,  
\ba
S_\text{IIB}&=&S_\text{IIB}^0+\gamma S_\text{IIB}^{1} +{\mathcal O}(\gamma^2) \, ,
\ea
where the ${\mathcal O}(\gamma^0)$ action has the usual form
\ba
S_\text{IIB}^0&=&\frac{1}{2\kappa_{10}}\int d^{10}x\sqrt{-g}\lk R_{10}-\frac{1}{2}(\partial \phi)^2-\frac{1}{4.5!}(F_5)^2 \, \rk 
\ea
and the term $S_\text{IIB}^{1}$ reads \cite{Myers:2008yi,Paulos:2008tn}
\ba \label{actioncorr}
S_\text{IIB}^{1}&=&\frac{L^6}{2 \kappa_{10}^2}\int d^{10}x\sqrt{-g} e^{\frac{-3}{2}}\phi \Big(C+\mathcal{T}\Big)^4 \, .
\ea
In these functions, $\kappa_{10}$ is the ten-dimensional gravitational constant, $R_{10}$ the Ricci scalar, $\phi$ the dilaton, $F_5$ the five-form field strength, and $C$ the Weyl tensor. The tensor ${\mathcal T}$ finally reads
\be
\mathcal{T}_{abcdef}=i\nabla_a F^+_{bcdef}+\frac{1}{16}\( F^+_{abcmn}F_{def}^{+\;\;mn}-3F^+_{abfmn}F_{dec}^{+\;\;mn}\) \, ,
\ee
where the index triplets $\{a,b,c\}$ and $\{d,e,f\}$ have been first antisymmetrized with respect to permutations, and the two triplets then symmetrized with respect to the interchange $abc\leftrightarrow def$. For further details of the construction of this action, such as the precise definitions of the various terms in eq.~(\ref{actioncorr}), see e.g.~\cite{Hassanain:2011ce}.

From the above starting point, one can derive several useful results, such as the $\gamma$-corrected black hole metric \cite{Paulos:2008tn,Gubser:1998nz,Pawelczyk:1998pb}
\ba\label{AdSg}
ds^2 = \frac{r_h^2}{u} \, \left(-f(u) \,
K^2(u) \, dt^2 + d\vec{x}^2\right) 
+ \frac{1}{4 u^2
f(u)} \, P^2(u) \, du^2 + L^2(u) \, d\Omega_5^2\, ,
\ea
where $u\equiv r_h^2/r^2$, $f(u)\equiv 1-u^2$, and the different functions read
\ba
\!\! K(u) &=& e^{\gamma \, [a(u) + 4b(u)]}\, , \,\, P(u) = e^{\gamma \,
b(u)}\,, \,\, L(u) = e^{\gamma \, c(u)}\, , \nonumber \\
a(u) &=& -\frac{1625}{8} \, u^2 - 175 \, u^4 + \frac{10005}{16} \,
u^6 \, , \nonumber \\
b(u) &=& \frac{325}{8} \, u^2 + \frac{1075}{32} \, u^4
- \frac{4835}{32} \, u^6 \, , \nonumber \\
c(u) &=& \frac{15}{32} \, (1+u^2) \, u^4 \,,
\ea
or the $\gamma$-corrected relation between the horizon radius $r_h$ and the field theory temperature, $r_h=\pi T/(1+\frac{265\g}{16})$. In addition, one obtains corrections to the equations of motion (EoMs) for various bulk fields, needed in determining Green's functions for the corresponding boundary operators. 

For the sake of concreteness, let us review here the case of the transverse component of a bulk U(1) electric field $E_\perp(u)$, dual to an electromagnetic current operator on the field theory side (see \cite{Steineder:2013ana} for details). First, define
\ba
\Psi(u) &\equiv& \Sigma(u) E_\perp(u)\, , \quad
\Sigma(u)^{-1} = \left\{ \begin{array}{lr}
1/\sqrt{f(u)}+\gamma p(u) \;\; \mathrm{for\; the\; black\; hole\; metric}, \\
1  \;\;\;\mathrm{for\; empty\; AdS\; space},
\end{array}\right. \label{sigmadef}
\ea
where the function $p(u)$ reads (below $\hat{q} \equiv q/(2\pi T)$, $\hat\omega \equiv \omega/(2\pi T)$)
\ba
p(u)&=&\frac{u^2\big(11700 - u^2 \big[343897 -u(87539\,u +37760\,\hat{q}^2)\big]\big)}{288\sqrt{f_+(u)}} \, .
\ea
This allows one to reduce the problem of finding the classical bulk field to that of solving a simple Schr\"odinger-type system with the action 
\ba
S&=&-\frac{N_c^2r_h^2}{16\pi^2}\int_k \int \!{\rm d}u\, \bigg[\frac{1}{2}\Psi{\mathcal L}\Psi +\partial_u \Phi \bigg]\, ,
\ea
where $\Phi(u)\equiv \Psi'(u)\Psi(u)$ and the EoM for $\Psi(u)$ reads
\ba
\Psi''(u)&-&V(u)\Psi(u)=0\, , \label{EoM1} \\
V(u)&=&-\frac{1}{f_+^2}\bigg[\frac{u+\hw^2-\hq^2 f_+}{u}- \gamma\frac{f_+}{144} \Big(-11700 u + 2098482 u^3 - 4752055 u^5 \nonumber \\
&&+ 1838319 u^7+ 
 \hat{q}^2 (4770 + 11700 u^2 - 953781 u^4 + 1011173 u^6) \nonumber \\
 &&-  \hw^2(4770  + 28170 u^2 - 1199223 u^4)\Big) \bigg]\, . \label{V}
\ea
From here, one can e.g.~solve the ${\mathcal O}(\gamma)$ bulk field obeying an ingoing boundary condition at the horizon of the black hole and thereby obtain the corresponding retarded Green's function in thermal equilibrium, giving rise to a $\gamma$-corrected QNM spectrum. Equivalently, following the $\gamma=0$ work of \cite{Baier:2012tc,Baier:2012ax}, one can consider the simple model of holographic thermalization involving a gravitationally collapsing planar shell. Then one simply imposes the infalling boundary conditions at the center of AdS space and uses the Israel junction conditions at the shell to glue together the classical field solutions in the parts of spacetime on the two sides of the shell. For the cases of the electromagnetic current and energy momentum tensor correlators, this proceduce is performed to linear order in $\gamma$ in \cite{Steineder:2012si,Steineder:2013ana,Stricker:2013lma}.

\section{Quasinormal modes and off-equilibrium Green's functions}

Next, we take a brief look at a selected set of recent ${\mathcal O}(\gamma)$ results for quantities relevant to the physics of thermalization. First, we inspect the QNM spectrum corresponding to the electromagnetic current correlator in thermal equilibrium, and then study the corresponding spectral function in the falling shell setup.

\subsection{QNM analysis}

\begin{figure}
  \includegraphics[height=.25\textheight]{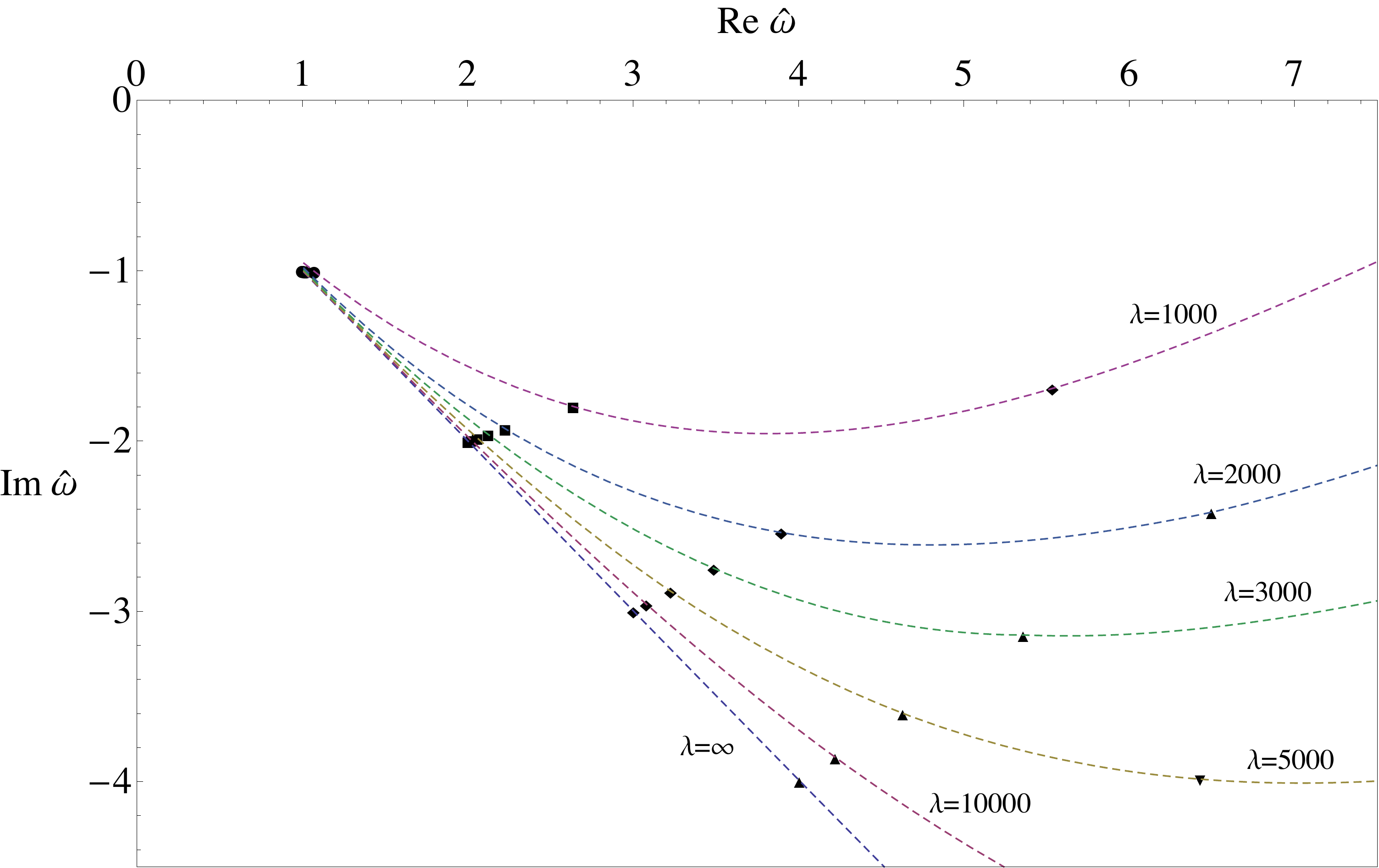}
  \caption{The QNM spectrum corresponding to the electromagnetic current operator in ${\mathcal N}=4$ SYM theory, diasplayed for several values of the 't Hooft coupling $\lambda$ \cite{Steineder:2013ana}. The dashed lines corresponding to various values of $\lambda$ are drawn here only to guide the eye.}
\end{figure}

The quasinormal mode spectrum corresponding to a given operator in ${\mathcal N}=4$ SYM theory is obtained from the pole structure of the retarded Green's function in question. It can be viewed as the strong coupling equivalent of a quasiparticle spectrum at weak coupling, even though for most operators studied the corresponding weak coupling correlators exhibit branch cut singularities rather than individual poles. The typical QNM spectrum has the form 
\ba
\omega_n(q)&=&M_n(q)-i\, \Gamma_n(q),
\ea
where the index $n \in {\mathcal Z}_+$ labels the individual poles, and $M_n(q)$ is interpreted as the energy and $\Gamma_n(q)$ as the width of an excitation with three-momentum $q$. For the case of the electromagnetic current operator at infinite 't Hooft coupling, the $q=0$ limit of the spectrum is known to take the simple form 
\ba
\omega_n(q=0)&=&2\pi Tn(\pm 1-i),
\ea
which is remarkably far from the limit of quasiparticles, characterized by widths parametrically smaller than the corresponding energies, $\Gamma_n \ll M_n$. 

In \cite{Steineder:2013ana}, the behavior of the above QNM spectrum was studied when reducing the value of $\lambda$ from infinity, i.e.~while taking the leading $\gamma$ correction into account. The result of this exercise, displayed in fig.~1, shows a clear pattern of the poles moving towards the real axis upon lowering $\lambda$, which is consistent with the spectrum moving towards a branch cut residing close to the real axis. A closer inspection of the $n$ dependence of the corrections also reveals that they grow as $n^4$, which is the naive expectation based on the form of the $\alpha'$ corrections to the SUGRA action.

\subsection{Retarded correlators in a thermalizing plasma}

Moving next on to quantities probing an out-of-equilibrium system, let us look at the behavior of the transverse component of the electromagnetic spectral function $\chi_\perp(\hw,\hq,u_s,\gamma)$ in the collapsing shell setup. In particular, it is interesting to inspect the behavior of this quantity as a function of the 't Hooft coupling $\gamma$, working in the quasistatic limit where the shell is taken to be a static object residing at some $u_s<1$. To this end, define a quantity
\be
\label{R2}
R_\perp(\hw,\hq,u_s,\gamma)=\frac{\chi_\perp(\hw,\hq,u_s,\gamma)-(\chi_\text{th})_\perp(\hw,\hq,\gamma)}{(\chi_\text{th})_\perp(\hw,\hq,\gamma)}\, ,
\ee
which clearly represents the relative deviation of the spectral function from its thermal limit. When viewed as a function of frequency, $R_\perp$ typically oscillates around zero and rapidly vanishes as the shell reaches the location of the final event horizon, $u_s\to 1$ \cite{Steineder:2013ana}. This can be interpreted as a clear signature of thermalization.

\begin{figure}
  \includegraphics[height=.2\textheight]{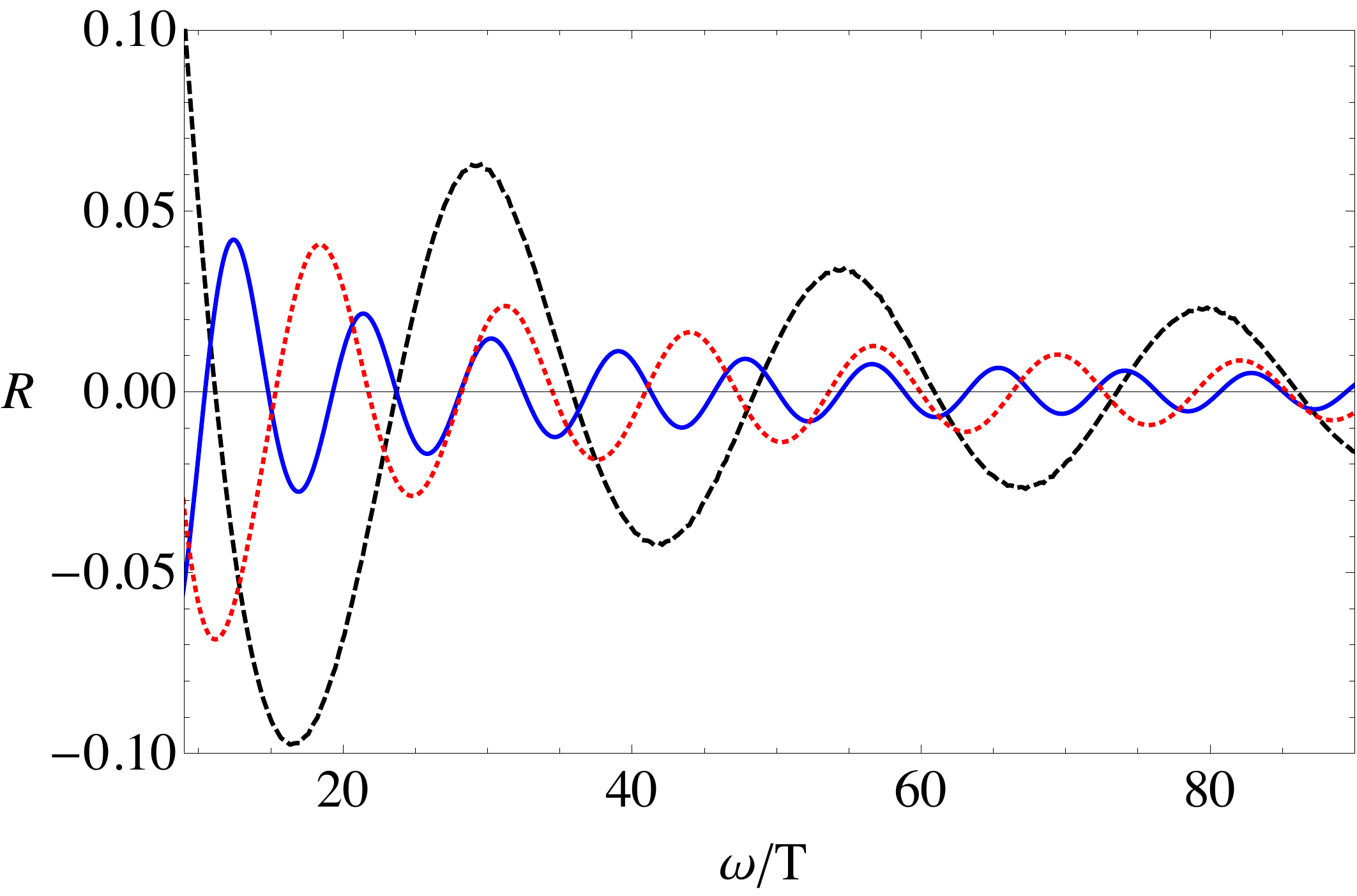}$\;\;\;\;\;$\includegraphics[height=.2\textheight]{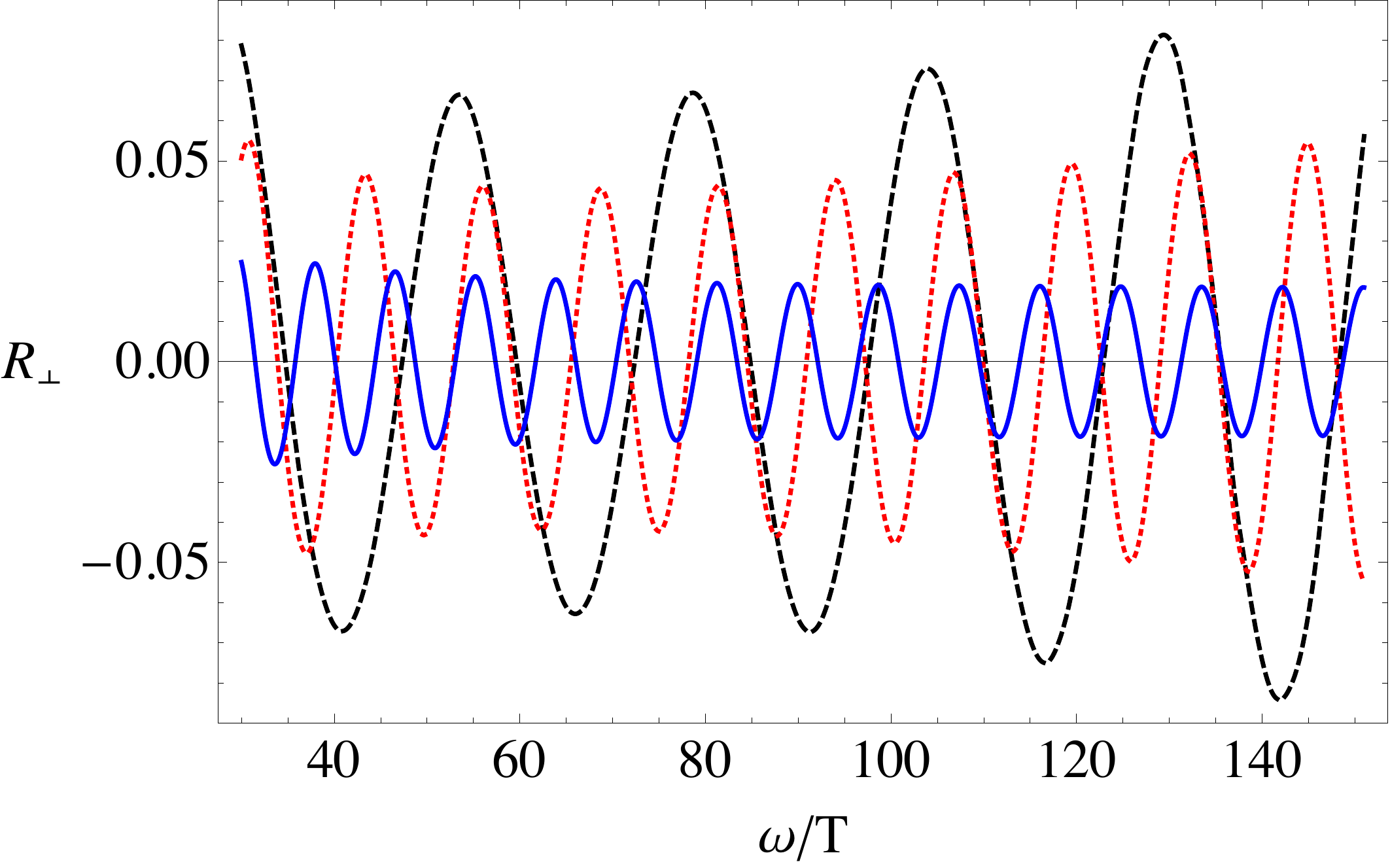}
  \caption{The relative deviation of the transverse spectral function from its thermal limit, evaluated for $q/\omega=$0 (blue solid curve), 0.8 (red dotted curve) and 1 (black dashed curve), and for $\lambda=\infty$ (left figure) and $\lambda=100$ (right figure) \cite{Steineder:2013ana}.}
\end{figure}

A very interesting feature of the relative deviation is the behavior of its amplitude as a function of $\omega$. At infinite coupling (fig.~2 left), the amplitude is seen to monotonously decrease with increasing $\omega$ --- a property typically interpreted as a sign of a top/down thermalization pattern, following also from purely causal arguments in the gravity picture. However, as soon as one lowers the value of $\lambda$ from infinity (fig.~2 right), it is seen that there exists a critical value of $\omega$, beyond which the amplitude starts to increase; this phenomenon is independent of the virtuality of the produced photons, even though the numerical value of $\omega_\text{crit}$ is rather sensitive to this parameter. It has been speculated in \cite{Steineder:2012si,Steineder:2013ana,Stricker:2013lma} that this behavior may reflect the onset of a qualitative change in the behavior of the system, indicating a move towards a bottom/up-type thermalization pattern at weak coupling.

\section{Future directions}

Taking studies of holographic thermalization away from the unphysical limits of infinite $\lambda$ and $N_c$ and superconformal invariance is a complicated but necessary goal, if one wants to be able to draw conclusions of quantitative significance for heavy ion or condensed matter physics. In the proceedings contribution at hand, I have reported on a set of recent calculations attempting to relax the first of these limits, namely study the thermalization process taking into account the leading $\alpha'$ corrections to the type IIB supergravity action and thereby including the first strong coupling corrections to the quantities under study \cite{Steineder:2012si,Steineder:2013ana,Stricker:2013lma}. As seen in the previous section, all results obtained in these studies point towards a weakening of the usual top/down thermalization pattern observed at infinite coupling, consistent with the slow transition of the system towards the expected weak coupling behavior.

So far, holographic thermalization studies have been generalized to non-infinite couplings only in a few very simple setups, typically involving a quasistatic gravitationally collapsing shell. In the future, the most important challenge will undoubtedly be the generalization of these calculations to more dynamical --- and more realistic --- models of holographic thermalization, first addressing simple quantities \cite{Keranen:2014zoa} and ultimately perhaps reaching the famous colliding shockwave scenario of \cite{Chesler:2010bi,Wu:2011yd,Heller:2012km}. A natural intermediate goal in this direction would be a more profound physical understanding of the current ${\mathcal O}(\alpha'^3)$ results, in particular inspecting why quantities related to thermalization appear to be so much more sensitive to strong coupling corrections than many others, such as bulk thermodynamic functions and transport coefficients (see e.g.~\cite{Gubser:1998nz,Buchel:2004di}).





\bibliographystyle{aipproc}   

\bibliography{confinement2014}

\IfFileExists{\jobname.bbl}{}
 {\typeout{}
  \typeout{******************************************}
  \typeout{** Please run "bibtex \jobname" to optain}
  \typeout{** the bibliography and then re-run LaTeX}
  \typeout{** twice to fix the references!}
  \typeout{******************************************}
  \typeout{}
 }

%
%
%
%
%

\end{document}